\RequirePackage[hyphens]{url}

\documentclass[twocolumn, final, 5p, 12pt]{elsarticle}

\usepackage{amssymb}

\usepackage{lineno}

\usepackage{booktabs}
\usepackage{graphicx}
\usepackage{multirow}
\usepackage{color}
\usepackage{url}
\usepackage{tabularx}
\usepackage{caption}
\usepackage{amsmath}
\usepackage[ruled]{algorithm2e}

\usepackage{subfigure}

\usepackage{seqsplit} 

\newcommand{\todo}[1]{\textcolor{red}{#1}}
\newcolumntype{L}[1]{>{\raggedright\let\newline\\\arraybackslash\hspace{0pt}}m{#1}}  

\journal{DFRWS EU 2023}

\begin{document}

\begin{frontmatter}



\title{Hamming Distributions of Popular Perceptual Hashing Techniques}


\author[inst1]{Sean McKeown}

            
 \affiliation[inst1]{organization={School of Computing, Engineering, and the Built Environment},
            addressline={Edinburgh Napier University}, 
             city={Edinburgh},
             country={UK}}

\author[inst1]{William J. Buchanan}


\begin{abstract}
Content-based file matching has been widely deployed for decades, largely for the detection of sources of copyright infringement, extremist materials, and abusive sexual media. 
Perceptual hashes, such as Microsoft’s PhotoDNA, are one automated mechanism for facilitating detection, allowing for machines to approximately match visual features of an image or video in a robust manner. However, there does not appear to be much public evaluation of such approaches, particularly when it comes to how effective they are against content-preserving modifications to media files.
In this paper we present a million-image scale evaluation of several perceptual hashing archetypes for popular algorithms (including Facebook’s PDQ, Apple’s Neuralhash, and the popular pHash library) against seven image variants. The focal point is the distribution of Hamming distance scores between both unrelated images and image variants to better understand the problems faced by each approach.
\end{abstract}



\begin{keyword}
Perceptual Hashing\sep Fuzzy Hashing\sep Hash Matching\sep CSAM\sep Image Forensics
\end{keyword}

\end{frontmatter}


\section{Introduction}
\label{sec:intro}






High-speed broadband has facilitated a transition from a text-based Internet to one with a media-heavy landscape, which is capitalised upon by platforms such as Facebook, YouTube, Instagram, TikTok, and others. Unfortunately, as with any technology, multimedia can be used for illegal and abusive purposes, necessitating that law enforcement agencies, social media platforms, cloud providers, and so on, engage in some form of content moderation. At Web scale the manual evaluation of multimedia becomes infeasible, resulting in the use of automatic detection technologies. 
Content-based detection is implemented for a variety of reasons, such as for the detection of extremist and violent material~\cite{microsoft_corporate_blog_christchurch_2019}, copyright infringing material~\cite{saadatpanah2020adversarial}, and the detection of Child Sexual Abuse Material (CSAM)~\cite{lee2020detecting}. The CSAM use-case is particularly important, as the Internet Watch Foundation reports a sharp increase in the detection and distribution of CSAM media - taking action against 250,000 websites in 2021 alone~\cite{the_internet_watch_foundation_iwf_2021}. Content matching technologies have already been deployed for many cloud-based services, backed by Microsoft's PhotoDNA~\cite{photodna} and Facebook's PDQ~\cite{facebook}, with Apple looking to introduce client-side device scanning~\cite{csam} to further combat the growing problem. 

For large-scale deployment, it is imperative that these detection technologies are well understood, both in terms of their strengths and weaknesses, in order to avoid improper configuration/deployment and incorrect actionable intelligence. The primary contribution of this work is an analysis of contemporary algorithms in still-image content-based perceptual hash matching, with analysis of their robustness in the face of content-preserving attacks, with experiments being conducted at the million-image scale.

\section{Background and Related Work}
\label{subsec:lit}



Content-based file matching can take a variety of forms, with the approach archetypes corresponding to: i) Exact matching with Cryptographic hashes; ii) Approximate similarity binary-level matching; and iii) Semantic approximate matching of file contents~\cite{breitinger_towards_2013}.

In each case, a hash digest is created which serves as a fingerprint/signature for the file, which can then be compared to the hashes of other files to detect identical or similar file content. Cryptographic approaches are commonly used in CSAM detection in digital forensics analysis, however, they are easily defeated by modifying a single-bit in a file. Approximate matching at the binary level is suitable for some file types, but is generally a poor choice for media files as simply changing encoding parameters will result in completely different binary content~\cite{farid_overview_2021}. We,  therefore, choose to focus on the semantic domain, which, for still-images, is the matching of visually similar content (as opposed to similarly sounding audio, or similar semantic content in text). For visual content, semantic domain matching is referred to as Perceptual Image Matching.

\subsection{Perceptual Image Matching}

Perceptual hashing~\cite{farid_overview_2021,hadmi_perceptual_2012} approaches are inspired by the domain of Content-Based Image Retrieval (CBIR)~\cite{rafiee_review_2010}, with the goal of providing the ability to measure visual image similarity and leverage this to return matches for visually identical, or similar, images. Ideally, this process should be robust to content-preserving attacks, such as adding noise, filters, cropping, and so on, but also to some degree, content modifying manipulations, perhaps involving the addition or removal of objects.



Hadmi et al.~\cite{hadmi_perceptual_2012} identified a typical pipeline for the creation of such robust hashes: \emph{(i)} Transformation, e.g., spatial transformations, or transformation to another domain, such as the frequency domain, with Discrete Cosine Transform (DCT) and Discrete Wavelet Transform (DWT) commonly being used; \emph{(ii)} Feature Vector Extraction; \emph{(iii)} Quantization and Feature Reduction; and \emph{(iv)} Compression or Encryption to generate a fixed-length hash. 


These hashes are then compared using a variety of similarity metrics (Hamming distance, Euclidean distance, Earthmover distance, L2 distance, and so on),  typically normalised to a value between zero and unity, with unity representing complete similarity. 


A wide array of features have been used as the basis for generating robust image signatures. Histograms and statistical information about the entire image may be used, providing a high-level representation which is insensitive to small, localised, changes in the image. This can take the form of colour histograms~\cite{swain_color_1991}, texture and edge histograms~\cite{manjunath_color_2001}, or frequency domain statistics~\cite{venkatesan_robust_2000}.  Properties of human vision may also be exploited, such as insensitivity to high-frequency changes in an image over a small area - a property which is exploited by JPEG compression~\cite{wallace_jpeg_1992}. Low frequency properties of an image may be used to derive a perceptual hash~\cite{fridrich_robust_1999, fridrich_robust_2000}, which provides robustness to compression artefacts and other content-preserving modifications. In a similar vein, coarse image representations may be used, such as low-resolution versions of the image, or the average colour value of sub-blocks in the image~\cite{steinebach_robust_2011}. Alternatively, invariant relationships in the image may be exploited, such as those found on radial lines projected out from the centre of the image~\cite{standaert_practical_2005}, which is particularly effective against re-scaled images.

It should also be noted that there is a distinction between hashes which are generated using a \emph{shallow}, heuristic, approach and those derived from a deep learning, data-driven, approach~\cite{dolhansky_adversarial_2020}. In the former case, hashes are generated in a deterministic manner from a set of operations (such as statistical features of the image), while the learning approaches will generate different hashes based on the dataset used to train the corresponding model.

\subsection{Evaluating Perceptual Image Matching Against Attack}

Despite the popularity and wide spread use of perceptual hashing, the robustness of these algorithms is not well understood for malicious and abusive content detection~\cite{hao_its_2021}, as adversaries can produce image variants that may hinder detection and identification.

One of the most popular, and easily available, perceptual hashing algorithms is pHash (DCT-based)~\cite{zauner2010implementation}. This hashing method is available in the Python ImageHash library~\cite{JohannesBuchner}, and is often evaluated together with the other algorithms in the library: ahash (average block colour); dhash (adjacent block colour); and whash (DWT-based)~\cite{dolhansky_adversarial_2020,hao_its_2021,hamadouche_comparative_2021,drmic_evaluating_2017,jain_adversarial_2022}. 

Hamadouche et al.~\cite{hamadouche_comparative_2021} studied the ImageHash algorithms together with SVD-Hash (Singular Value Decomposition) against a variety of image-wide filters, noise, and scaling for a small dataset of 800 images. While ahash typically produced the smallest distances between original and modification, pHash and dhash were the only algorithms to produce normally distributed Hamming distances - a property which will be discussed further in Section~\ref{subsecsec:criteria}. Jain et al.~\cite{jain_adversarial_2022} performed more sophisticated perturbation attacks against ImageHash algorithms and Facebook's DCT-based PDQ on over one million images. The authors noted that by using black-box attacks, it was possible to manipulate modified image distances to the extent that the False Positive Rate would become unacceptably large in all cases. Similar work by Hao et al.~\cite{hao_its_2021} for ImageHash and Blockhash (average block colour) allowed for large distances to be achieved for each algorithm using noise, cropping, rotation and scaling attacks.
Dmric et al.~\cite{drmic_evaluating_2017} evaluated the ImageHash library, with 1,480 images, against user-level modifications, such as resizing, rotation, adding borders, etc., together with social media post-processing tests. pHash generated the best F1 score (weighted combination of precision/recall) when aggregated across attacks, with social media manipulations disturbing performance less than the more direct manipulations.

McKeown and Russell~\cite{mckeown_fast_2019} explored the use of pHash and Blockhash for matching originals to thumbnail cache entries in Windows Vista, 7, and 10, noting that neither algorithm could achieve an acceptable False Positive Rate to False Negative Rate trade-off for forensics purposes. However, the combination of both algorithms may be sufficient. The authors also noted particular weaknesses in the algorithms, such as fractal/patterned images for pHash and solid colour background/smooth gradients for Blockhash.

Looking more widely, Zauner~\cite{zauner2010implementation} and Breitinger et al.~\cite{breitinger_towards_2013} compared DCT, Block Mean Based, Radial (e.g. Radon projection) and Marr-Hildreth (MH) operator hashes. Particular attention was paid to the strengths and weaknesses of each algorithm for each modification attack. JPEG compression did not have much effect on any tested algorithm, while resizing only particularly affected MH (most algorithms downscale images in pre-processing). Mirroring, rotating, and cropping all have much larger effects on performance across attacks, except when algorithms build in specific handling (such as for rotation in rHash~\cite{steinebach_robust_2011}).

Machine learning approaches have also been shown to be susceptible to trivial content-preserving manipulations, with Struppek et al.~\cite{struppek_learning_2022} demonstrating that Apple's Neuralhash is not robust to gradient and familiar transformation-based attacks, with hash collision attacks also being possible. Dohlansky et al.~\cite{dolhansky_adversarial_2020} found similar results for AlexNet, ResNet, and EfficientNet, although the authors noted that cross-attacks between shallow and deep approaches do not work well, such that combining algorithms from each class may lead to additional robustness against attack.

\section{Methodology}
\label{sec:methodology}

While there is a body of existing work in the evaluation of perceptual hashing, there is generally a focus on aggregating statistics across multiple attack types, or in generating specific attack scenarios, rather than understanding the behavioural properties of the approaches against common user-level attacks. With a similar approach to Hamadouche et al.~\cite{hamadouche_comparative_2021}, we set out to understand the Hamming distance distribution of perceptual hashing algorithms in order to better quantify behaviour for not only the aggregate/average case, but also the best and worst case scenarios. To achieve this, we make use of the Flickr 1 Million dataset~\cite{noauthor_mirflickr_nodate}, removing SHA256 hash duplicates, which allows us to scale experiments to large numbers of natural images. The original hashes, Hamming scores, and additional statistical data used in this paper are available online with DOI: \url{10.5281/zenodo.7426035}.

\subsection{Selection of Hashing Algorithms}

Our perceptual hash selection takes into account a wide range of feature extraction techniques, but also focuses on popular algorithms. The chosen hash algorithms are listed in Table~\ref{table:hashes}. The list covers: frequency transforms; colour histograms; block mean colour; and deep learning-derived approaches. Aside from additional deep learning approaches, one notable omission of the shallow approach to hashing in this testing is Microsoft's PhotoDNA~\cite{photodna} as it is a controlled technology and not readily available. While such evaluations may already be available to those with access, it would be beneficial for future work to make evaluations open-source, particularly as it is the de-facto standard for the perceptual detection of CSAM~\cite{farid2018reining}. The details of its implementation are not widely detailed, however it appears that PhotoDNA uses an edge-based method~\cite{prokos_2021_squint}, such that it is unclear what particular weaknesses it may share with the algorithms presented in this paper.
Another possible algorithm class to include in future testing is the Scale Invariant Feature Transform (SIFT), a feature-point based approach, which is more commonly used for image forgery detection~\cite{bourouis2020recent}. ForBild/rHash~\cite{steinebach_robust_2011} was also considered, though it was excluded early on as the sourced implementation generated many hash collisions due to black backgrounds, or similar colour compositions.

\begin{table*}[t]
\footnotesize
\centering
\begin{tabular}{|L{2cm}|L{15.5cm}|}
\hline
\textbf{Blockhash} & An implementation of the Block Mean Based approach~\cite{yang_block_2006}. It can be found, written in C, on the Commons Machinery Github~\cite{commonsmachinery_contribute_2018}. Generates 256-bit hashes. \\ \hline
\textbf{ColourHash} & Part of the  Python ImageHash Library~\cite{JohannesBuchner}. Images are matched based on colour distributions. Generates 44-bit hashes. \\ \hline
\textbf{NeuralHash (Apple)} & Apple's CSAM hashing scheme. This is a machine-learned approached, with the neuralhash\_128x96\_seed1.dat model being   extracted from an \url{iPhone14,5_15.3.1_19D52} IPSW firmware image. The model extraction process and code for generating hashes, can be found on Github~\cite{ygvar_appleneuralhash2onnx_2022}. Generates 96-bit hashes. \\ \hline
\textbf{PDQ (Facebook)} & Facebook's   improved version of the pHash (DCT) algorithm, with optimisations for   downsampling, and larger hash output size by default (for better Web-scale accuracy). The Python implementation - available on Facebook's Threat Exchange Github~\cite{facebook} - was used. Generates 256-bit hashes. \\ \hline
\textbf{pHash} & DCT-based, as implemented in the ImageHash Python library~\cite{JohannesBuchner}. Generates 64-bit hashes. \\ \hline
\textbf{Wavehash} & A hash based on the Discrete Wavelet Transform (DWT) (as opposed to the DCT based   approaches above). Also part of the Python ImageHash library~\cite{JohannesBuchner}. Generates 64-bit hashes. \\ \hline
\end{tabular}
\caption{Perceptual Hashes used in experiments.}
\label{table:hashes}
\end{table*}


\subsection{Image Modifications}

A set of six image modifications were chosen as \emph{attacks} against the perceptual hashing algorithms. Each modification was implemented using the Python PIL library~\cite{ab_pythonware_pil_nodate}. This means that all images were modified using the same tool, introducing some limitations in terms of the diversity of the images produced, but we do not expect this to have a significant effect on the outcomes.

\begin{figure}
\begin{center}
	\includegraphics[width=0.85\linewidth]{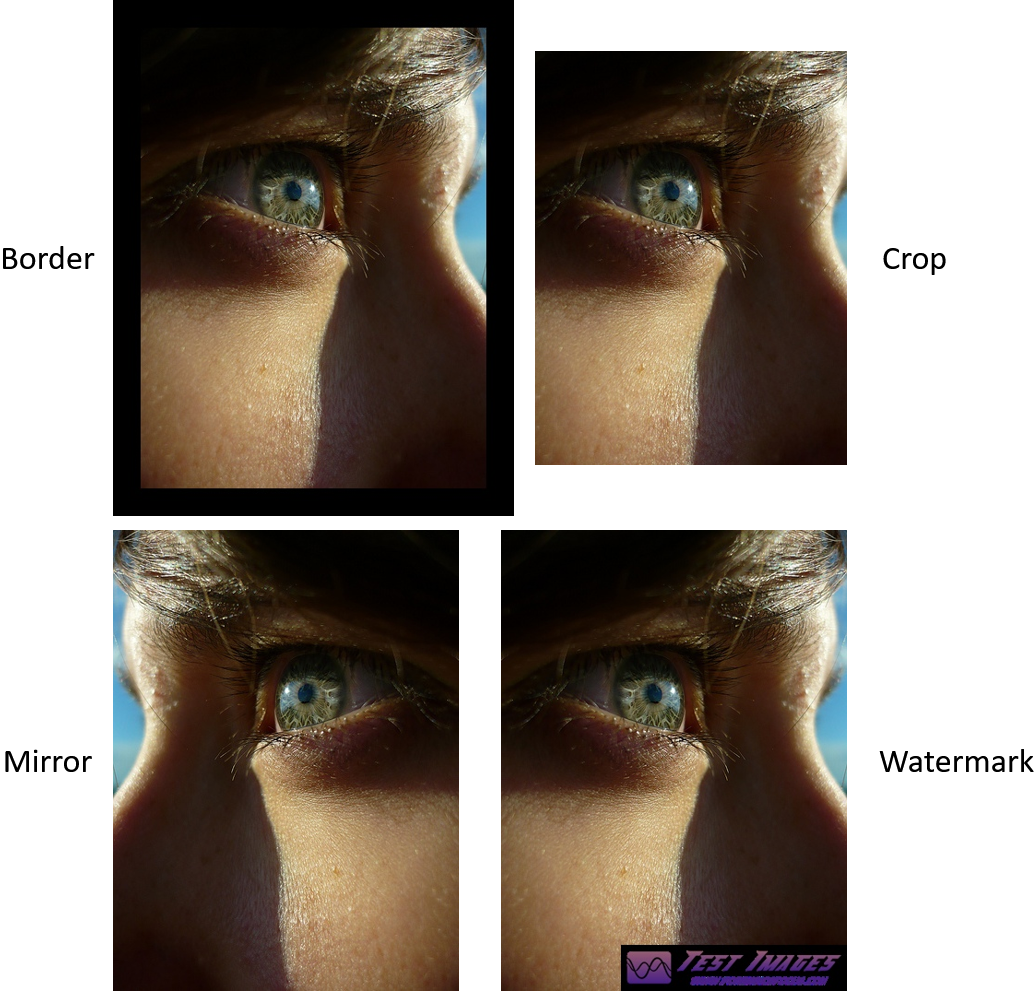}
\end{center}
\caption{Visibly distinct modification examples for 50.jpg in the dataset.}
	\label{fig:mods_example}
\end{figure}

To reduce the total processing time across all hash and modification permutations, a random subset of 250,000 images was selected for the creation of modifications. Table~\ref{table:mods} describes the chosen modifications, which are intended to reflect the low-barrier to entry, unsophisticated, content-preserving black-box attacks that a typical user may employ which do not compromise viewability (such as for evading copyright detection~\cite{jabade_modelling_2016}), with the addition of Windows 10 generated thumbnails as a realistic downscaling exemplar. Examples of the visually distinct modifications are provided in Figure~\ref{fig:mods_example}.

\begin{table*}[t]
\footnotesize
\centering
\begin{tabular}{|L{2cm}|L{15.5cm}|}
\hline
\textbf{Border (30px)}         & Add a fixed size 30-pixel black border to the outside of the image, extending it, rather than overwriting any of the existing content.                                                                                                                                                                     \\ \hline
\textbf{Compression (Q30)}   & Reduce the JPEG image quality to 30\%, scaling down the default quantization   tables. This simulates lower-quality images (but not lower resolution), which   can affect perceptual hashing performance. All other modifications passed the   quantization tables through from the original without manipulation, which is on average  96\%  quality for images in the Flickr 1   Million dataset. \\ \hline
\textbf{Crop (5\%) }           & Remove 5\% of   the image from the top, left, right, and bottom of the image, in effect   reducing the overall pixel count to 81\% of the original ($0.9 height \times 0.9 width$ ).                                                                                                                                                                                                                                                     \\ \hline
\textbf{Mirror (x-axis)}    & Flip the   image on its x-axis, preserving viewability, but generating significant   pixel/binary level changes.                                                                                                                                                                                                                                                                  \\ \hline
\textbf{Scale (1.5x)}        & Scale the image up to 150\% the size of the original, using the PIL~\cite{ab_pythonware_pil_nodate}  \texttt{Image.resize} function.                                                                                                                                                                                                                                                                                                            \\ \hline
\textbf{Thumbs96 (Windows)} & Generate   legitimate Windows 10 $96 \times 96$ pixel thumbnails as per the methodology in   McKeown and Russel~\cite{mckeown_fast_2019}, which generate a larger Hamming distance in this cited work for pHash and Blockhash than their $256 \times 256$   thumbnail counterpart.).                                                 \\ \hline
\textbf{Watermark }           & Add a   watermark to the bottom right of the image. The watermark consists of a logo,   text and a URL. The watermark was scaled to 10\% of the image height, with a   minimum of 40 pixels.                                                                                                                                                                                                                                       \\ \hline
\end{tabular}
\caption{Image modifications applied to each image in the dataset.}
\label{table:mods}
\end{table*}

\subsection{Evaluation Process and Criteria}
\label{subsecsec:criteria}

There are several things to consider when evaluating perceptual hashing systems. The first is the relative discrimination behaviour between unrelated images (inter-image). When measuring the distance between two hashes (usually via normalised Hamming distance), unrelated images should be essentially tossing a coin for each bit, resulting in a distance of around 0.5, on average (mean and median). Additionally, we should expect that very close hashes, and particularly exact matches, should not occur in large groups, and the occurrences should be reasonable and expected, that is, highly similar images. 

In order to characterise \textbf{inter-score } behaviour for the original (unmodified) images, each image in the Flickr 1 Million dataset was compared to a random selection of 50 other images, for a total of approximately 50 million comparisons\footnote{There are a possible total of 500 billion comparisons ($\frac{1million \times 1million}{2}$) for 1 million images, which is unnecessary to acquire the understanding we need here.} to generate statistics from. To better understand the impact of the various image modifications on the behaviour of each hash, inter-scores were also analysed for each of the seven modifications, resulting in 12.5 million additional samples per modification for each hashing algorithm. Results for these comparisons (via normalised Hamming distances) are reported in Section~\ref{subsec:interscore}.

The second main characteristic is the performance of a perceptual hashing algorithm when comparing an image to its variant. This second image could be a pixel perfect copy, a re-encoded version of the image into a different file format, a thumbnail, cropped manipulation, etc. Ideally the perceptual hashing algorithm generates the same, or similar, hash for each of these images, such that the Hamming distance is zero, or very small. The distance distributions of inter- and intra-scores for each algorithm then allow for a wider understanding of potential false positive and false negative rates for a given threshold distance, as discussed in prior work~\cite{hao_its_2021,mckeown_fast_2019}.

We explore the \textbf{intra-score} characteristics of the various algorithms and image modifications in Section~\ref{subsec:intrascore}. Each modified image in the dataset was hashed and compared to its original version, generating 250,000 Hamming distance comparisons for each hash algorithm/modification pair. Results present not only measures of distribution (range, mean, median, and standard deviation), but also exact hash match percentages to demonstrate particularly strong matching performance.
\section{Findings}
\label{sec:findings}

\subsection{Inter-score (Different Images)}
\label{subsec:interscore}

The findings for inter-image distributions (i.e., between unrelated images) are presented first, with Section~\ref{subsubsec:originaltooriginal} describing each algorithm's behaviour for the original Flickr 1 Million image set. This essentially acts as a baseline for each algorithm's behaviour when images are not expected to match. Section~\ref{subsubsec:modifiedtomodified} compares images in each modification class (e.g., cropped to cropped) in order to determine if there is any bias introduced by the modifications themselves, even if the images are still assumed to be unrelated. This is followed by Section~\ref{subsec:intrascore} which moves on to examine the impact that the modifications have when comparing the same image to versions of itself, which provides a different window into the biases and difficulties produced by the modified versions of the images.

\begin{table*}[h!]
    \centering
    \begin{tabular}{lrrrcr}
    \hline
     \textbf{hash algorithm}       &   \textbf{mean} &   \textbf{median} &   \textbf{stdev} & \textbf{range}   \\
    \hline
     blockhash  & 0.4923 &   0.4922 &  0.0785 & 0.0078--1.0000 \\
     colourhash & \todo{0.1601} &   \todo{0.1591} &  0.0547 & 0.0000--0.3636 \\
     neuralhash & 0.4973 &   0.5000 &  0.0600 & 0.0521--0.7917  \\
     pdq        & 0.5000 &   0.5000 &  0.0321 & 0.0391--0.6953  \\
     phash      & 0.4904 &   0.5000 &  0.0649 & 0.0938--0.8438  \\
     wavehash   & 0.4854 &   0.5000 &  0.1241 & 0.0000--1.0000  \\
    \hline
    \end{tabular}
    \caption[Inter-score normalised Hamming distances between random images.]{\textbf{Inter-score }normalised Hamming distances between random images in the Flickr 1 Million dataset. Each image was compared to 50 random images. The ideal value is 0.5 distance to make best use of the Hamming space.}
    \label{table:interoriginal}
\end{table*}

\begin{figure*}[!]
    \centering
    \subfigure{\includegraphics[width=0.42\textwidth]{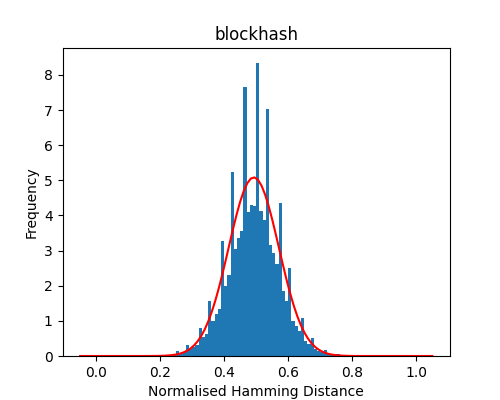}} 
    \subfigure{\includegraphics[width=0.42\textwidth]{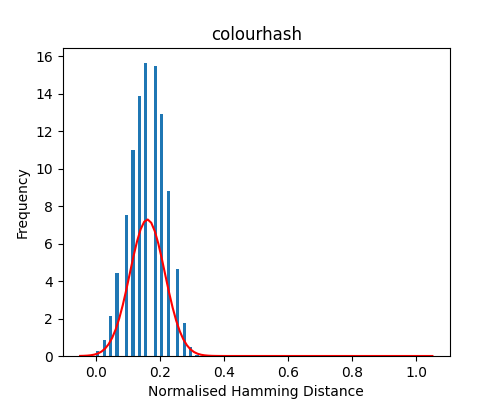}} 
    \subfigure{\includegraphics[width=0.42\textwidth]{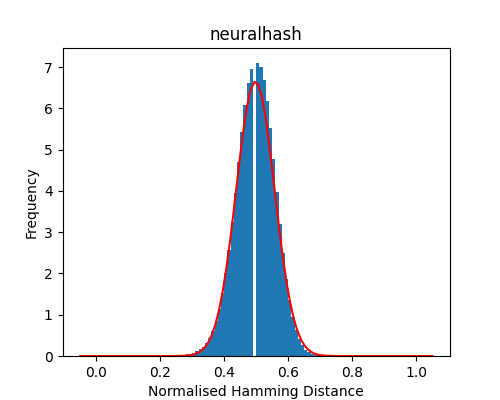}} 
    \subfigure{\includegraphics[width=0.42\textwidth]{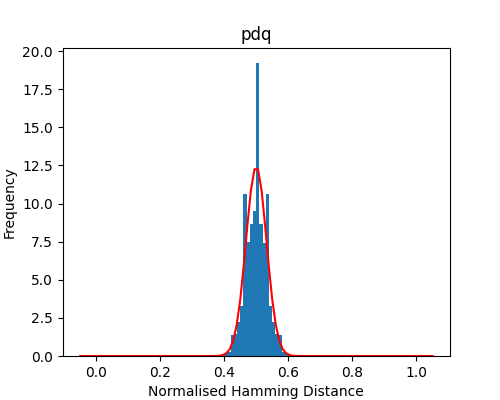}} 
    \subfigure{\includegraphics[width=0.42\textwidth]{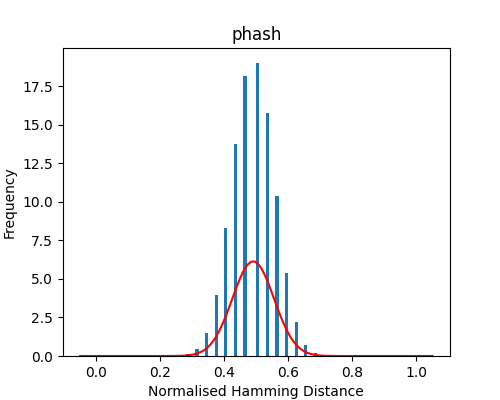}} 
    \subfigure{\includegraphics[width=0.42\textwidth]{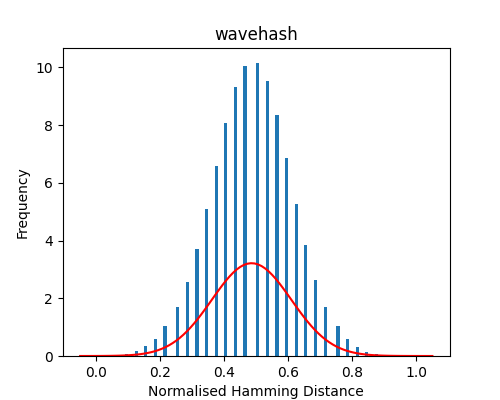}} 
    \caption[Inter-score normalised Hamming distance distributions.]{Normalised Hamming \textbf{inter-scores} with Normal Probability Density Function (NPDF) (50 million pair-wise samples) for the original Flickr 1 Million dataset. a) Blockhash, b) Colourhash, C) NeuralHash, d) PDQ, e) pHash, f) Wavehash}
    \label{fig:intraoriginals}
\end{figure*}
\subsubsection{Original to Original}
\label{subsubsec:originaltooriginal}

The results for the inter-score comparisons between original images in the dataset, for each hash, are presented in Table~\ref{table:interoriginal}, with distributions depicted in Figure~\ref{fig:intraoriginals}.

While almost all of the algorithms do well to centre inter-image scores around 0.5, which is clearly a design goal (particularly achieved by PDQ), Colourhash behaves quite differently. Colourhash has much lower scores for unrelated images, which is not necessarily detrimental in itself, however, it does mean that it makes less uniform use of the Hamming space. This will result in a lack of diversity, and more hash collisions due to chance.

When comparing the equivalence classes (distance of zero), the shortcomings of Colourhash are clear, with three equivalence classes containing over 20,000 images each, with many more classes with members in the thousands, or tens of thousands, making it functionality indiscriminate. Wavehash, despite its reasonable mean/median scores, produced a single class of 5,527 items, and several classes in the hundreds, again making it a poor discriminator at scale. Of the remaining algorithms, the only one which had a class size higher than size 5 is Blockhash, with aberrant classes of size 52 and 54.

In terms of their distributions, the algorithms which use lower hash sizes by default generate sparse distributions, with some distance intervals being unused, essentially quantizing the output across the Hamming space. This is particularly clear for Colourhash, pHash and Wavehash, where the sparseness creates a large mismatch with the Normal Probability Density Function (NPDF) in Figure~\ref{fig:intraoriginals}. NeuralHash has an intermediate hash length, but fits the NPDF much better, despite a notable gap in the centre of the distribution. The spikes produced by Blockhash and PDQ, which both use large hashes, are therefore likely algorithmic, making certain hash codes and distances more likely to occur by chance.

Of the hashes which conform to 0.5 as the median/mean, Wavehash has the widest distribution (highest standard deviation), with tails making it reasonably likely that it generates a zero distance match by chance at any reasonable scale, as is evident from the previously noted equivalence classes of over 20,000 items. This makes it unsuitable at large scale, even prior to attack. PDQ, on the other hand, has a very tight distribution, which is likely by design for scaling purposes to greatly reduce false positives. 

\subsubsection{Modified to Modified}
\label{subsubsec:modifiedtomodified}

From the equivalence class analysis alone, the Colourhash and Wavehash algorithms should be discounted as potential solutions. However, Wavehash remains in the following tables to provide insight into how a non-DCT based frequency domain transform is impacted by the various modifications.

Table~\ref{table:intermod} represents inter-score distances for each algorithm where the images in each modification class are compared to each other (e.g. cropped-to-cropped). The perceptual hashes largely remain stable across the modified image comparisons, though adding in repeated content in the form of a border has had a larger than expected impact for some hashes. Blockhash, pHash and Wavehash all have sizable reductions in mean distance between unrelated images as a result of the shared border. This means that they are more likely to match images with borders to each other, regardless of the content residing within the border. NeuralHash appears to be the only algorithm which is not particularly affected by the addition of a border. The watermark has a similar effect, though it is not nearly as pronounced.

\begin{table*}[hp]
    \centering
    \begin{tabular}{llrrrcr}
    \hline
     \textbf{hash algorithm}       & \textbf{modification}   &   \textbf{mean} &   \textbf{median} &   \textbf{stdev} & \textbf{range}          \\
    \hline
     blockhash  & border         & \todo{0.3284} &   \todo{0.3281} &  0.0549 & 0.0391--0.8047  \\
     blockhash  & compression    & 0.4922 &   0.4922 &  0.0782 & 0.0391--0.9805  \\
     blockhash  & crop           & 0.4941 &   0.4922 &  0.0761 & 0.0234--0.9766       \\
     blockhash  & mirror         & 0.4921 &   0.4922 &  0.0784 & 0.0312--0.9609 \\
     blockhash  & scale          & 0.4922 &   0.4922 &  0.0783 & 0.0391--0.9492 \\
     blockhash  & thumbs96       & 0.4921 &   0.4922 &  0.0787 & 0.0234--0.9609 \\
     blockhash  & watermark      & 0.4789 &   0.4766 &  0.0758 & 0.0391--0.9531 \\
    \hline
     neuralhash & border         & 0.4969 &   0.5000 &  0.0601 & 0.0000--0.7812 \\
     neuralhash & compression    & 0.4974 &   0.5000 &  0.0600 & 0.0729--0.7812 \\
     neuralhash & crop           & 0.4976 &   0.5000 &  0.0599 & 0.0833--0.7812 \\
     neuralhash & mirror         & 0.4974 &   0.5000 &  0.0600 & 0.0521--0.8021 \\
     neuralhash & scale          & 0.4973 &   0.5000 &  0.0600 & 0.0417--0.7812 \\
     neuralhash & thumbs96       & 0.4969 &   0.5000 &  0.0602 & 0.0312--0.7812 \\
     neuralhash & watermark      & 0.4969 &   0.5000 &  0.0602 & 0.0521--0.7812 \\
    \hline
     pdq        & border         & 0.4794 &   0.4766 &  0.0363 & 0.2109--0.6719 \\
     pdq        & compression    & 0.5000 &   0.5000 &  0.0321 & 0.2500--0.6797 \\
     pdq        & crop           & 0.5000 &   0.5000 &  0.0318 & 0.2891--0.7109 \\
     pdq        & mirror         & 0.5000 &   0.5000 &  0.0321 & 0.2969--0.6797 \\
     pdq        & scale          & 0.5000 &   0.5000 &  0.0321 & 0.2812--0.7188\\
     pdq        & thumbs96       & 0.5000 &   0.5000 &  0.0321 & 0.2812--0.7031 \\
     pdq        & watermark      & 0.4941 &   0.4922 &  0.0333 & 0.2031--0.7344 \\
    \hline
     phash      & border         & \todo{0.4283} &   \todo{0.4375} &  0.0727 & 0.0625--0.8438 \\
     phash      & compression    & 0.4904 &   0.5000 &  0.0649 & 0.1562--0.8438 \\
     phash      & crop           & 0.4907 &   0.5000 &  0.0641 & 0.0625--0.8125 \\
     phash      & mirror         & 0.4904 &   0.5000 &  0.0649 & 0.1562--0.8438 \\
     phash      & scale          & 0.4904 &   0.5000 &  0.0649 & 0.1250--0.8438 \\
     phash      & thumbs96       & 0.4904 &   0.5000 &  0.0650 & 0.1250--0.8125 \\
     phash      & watermark      & 0.4704 &   0.4688 &  0.0674 & 0.0000--0.8125 \\
     \hline
     wavehash   & border         & \todo{0.3013} &   \todo{0.2812} &  0.1167 & 0.0000--0.9688 \\
     wavehash   & compression    & 0.4854 &   0.5000 &  0.1240 & 0.0000--1.0000 \\
     wavehash   & crop           & 0.4869 &   0.5000 &  0.1212 & 0.0000--1.0000 \\
     wavehash   & mirror         & 0.4854 &   0.5000 &  0.1239 & 0.0000--1.0000 \\
     wavehash   & scale          & 0.4854 &   0.5000 &  0.1240 & 0.0000--1.0000 \\
     wavehash   & thumbs96       & 0.4854 &   0.5000 &  0.1241 & 0.0000--1.0000 \\
     wavehash   & watermark      & 0.4631 &   0.4688 &  0.1174 & 0.0000--1.0000 \\
    \hline
    \end{tabular}
    \caption[Inter-score normalised Hamming distances within each image modification.]{\textbf{Inter-score} normalised Hamming distances between random images within each modification category for the Flickr 1 Million dataset. Each image was compared to 50 random images (e.g. each cropped image was compared to 50 random cropped images for the same algorithm). The ideal mean/median is a distance of 0.5, demonstrating that the modification does not bias the algorithm.}
    \label{table:intermod}
\end{table*}

\begin{table*}[hp]
    \centering
\begin{tabular}{llrrrcr}
\hline
 \textbf{hash algorithm}       & \textbf{modification}   &   \textbf{mean} &   \textbf{median} &   \textbf{stdev} & \textbf{range}          &   \% exact matches \\
\hline
 blockhash  & border         & 0.2783 &   0.2734 &  0.0851 & 0.0000--\todo{0.7188} &      0.0000 \\
 blockhash  & compression    & \textbf{0.0095} &   \textbf{0.0078} &  0.0142 & 0.0000--\todo{0.4492} &     36.4310 \\
 blockhash  & crop           & 0.1668 &   0.1641 &  0.0610 & 0.0000--\todo{0.7344} &      0.0180 \\
 blockhash  & mirror         & \todo{0.4450} &   \todo{0.4531} &  0.1410 & 0.0000--\todo{1.0000} &      0.0460 \\
 blockhash  & scale          & \textbf{0.0013} &   \textbf{0.0000} &  0.0040 & 0.0000--0.1719 &     \textbf{85.4360} \\
 blockhash  & thumbs96       & \textbf{0.0254 }&   \textbf{0.0234} &  0.0179 & 0.0000--\todo{0.4961} &      5.5890 \\
 blockhash  & watermark      & \textbf{0.0504} &   \textbf{0.0469 }&  0.0305 & 0.0000--0.3672 &      2.9450 \\
 \hline
 neuralhash & border         & 0.0763 &   0.0729 &  0.0381 & 0.0000--\todo{0.5312} &      0.3670 \\
 neuralhash & compression    & \textbf{0.0082} &   0.0104 &  \textbf{0.0105} & 0.0000--0.3333 &     49.1630 \\
 neuralhash & crop           & \textbf{0.0605} &   \textbf{0.0521} &  0.0332 & 0.0000--\todo{0.4688} &      1.1540 \\
 neuralhash & mirror         & 0.2823 &   0.2812 &  0.1251 & 0.0000--\todo{0.6979} &      0.0960 \\
 neuralhash & scale          & \textbf{0.0036} &   \textbf{0.0000} &  0.0069 & 0.0000--0.1458 &     \textbf{73.6020} \\
 neuralhash & thumbs96       & 0.0809 &   0.0729 &  0.0445 & 0.0000--\todo{0.4792} &      0.9690 \\
 neuralhash & watermark      & \textbf{0.0551 }&   \textbf{0.0417 }&  0.0424 & 0.0000--\todo{0.5729} &      3.6560 \\
  \hline
 pdq        & border         & \todo{0.3949} &   \todo{0.3984} &  0.0599 & 0.0547--\todo{0.7422} &      0.0000 \\
 pdq        & compression    & \textbf{0.0094} &   \textbf{0.0078 }&  0.0091 & 0.0000--\todo{0.4453} &     23.4350 \\
 pdq        & crop           & 0.3255 &   0.3281 &  0.0564 & 0.0234--\todo{0.6719} &      0.0000 \\
 pdq        & mirror         & \todo{0.4975} &   \todo{0.5000} &  0.0226 & 0.0000--\todo{0.9844} &      0.0020 \\
 pdq        & scale          & \textbf{0.0237} &   \textbf{0.0234} &  0.0132 & 0.0000--\todo{0.4609} &      1.6320 \\
 pdq        & thumbs96       & 0.0721 &   0.0703 &  0.0281 & 0.0000--\todo{0.4844} &      0.0060 \\
 pdq        & watermark      & 0.1029 &   0.0938 &  0.0489 & 0.0000--\todo{0.5469} &      0.0020 \\
  \hline
 phash      & border         & 0.2656 &   0.2500 &  0.0745 & 0.0000--\todo{0.6562} &      0.0000 \\
 phash      & compression    & \textbf{0.0053 }&   \textbf{0.0000 }&  0.0138 & 0.0000--\todo{0.4688} &     \textbf{83.9040} \\
 phash      & crop           & 0.1686 &   0.1562 &  0.0586 & 0.0000--\todo{0.6562} &      0.0430 \\
 phash      & mirror         & \todo{0.4904} &   \todo{0.5000} &  0.0339 & 0.1562--\todo{0.6875} &      0.0000 \\
 phash      & scale          & \textbf{0.0020} &   \textbf{0.0000 }&  0.0091 & 0.0000--\todo{0.4688} &     \textbf{94.0050} \\
 phash      & thumbs96       & \textbf{0.0245} &   \textbf{0.0312 }&  0.0250 & 0.0000--\todo{0.5938} &     39.0830 \\
 phash      & watermark      & 0.1227 &   0.1250 &  0.0824 & 0.0000--\todo{0.6875} &      4.5350 \\
  \hline
 wavehash   & border         & 0.2744 &   0.2500 &  0.1333 & 0.0000--\todo{0.8750} &      0.2570 \\
 wavehash   & compression    & \textbf{0.0029} &  \textbf{0.0000} &  0.0106 & 0.0000--\todo{0.5312} &     \textbf{91.4070} \\
 wavehash   & crop           & 0.1049 &   0.0938 &  0.0649 & 0.0000--\todo{0.8750} &      3.5740 \\
 wavehash   & mirror         & 0.3474 &   0.3125 &  0.1698 & 0.0000--\todo{1.0000} &      1.2670 \\
 wavehash   & scale          & \textbf{0.0007} &   \textbf{0.0000} &  0.0052 & 0.0000--\todo{0.8438} &     \textbf{97.9830} \\
 wavehash   & thumbs96       & \textbf{0.0158 }&   \textbf{0.0000 }&  0.0214 & 0.0000--\todo{0.5938} &     57.3560 \\
 wavehash   & watermark      & \textbf{0.0451} &   \textbf{0.0312 }&  0.0483 & 0.0000--\todo{0.4688} &     39.6030 \\
\hline
\end{tabular}
    \caption[Intra-score normalised Hamming distances between original and modified version.]{\textbf{Intra-score} normalised Hamming distances between the original image and various modifications of the original for the Flickr 1 Million dataset. The ideal distance value is 0, which is reflected in the `\% matches' column. Bolded items are very good performances, while items in red are bad, to the point of being as bad, or worse, than selecting a random, unrelated, image.}
    \label{table:intramod}
\end{table*}

\subsection{Intra-score (Versions of the Same Image)}
\label{subsec:intrascore}

Intra-score values, comparing the original image to its modifications, are depicted in Table~\ref{table:intramod}. These scores give us an opportunity to explore how well the various techniques can deal with attempts to circumvent content-based detection.

All algorithms stumble into very high Hamming distances at some point, represented by the high maximum distance values, which simply suggests that there is at least one image in the dataset that troubles each algorithm for almost every modification. In some cases, the distance approaches one, which may actually be worthy of considering to be a match, as given a normal distribution around 0.5, a distance of one is just as likely as a distance of zero. These outliers are often hidden when considering aggregate behaviour, and they do not always occur, but particularly troublesome ranges are present for \emph{Blockhash-Mirror}, \emph{PDQ-Border}, \emph{PDQ-Mirror}, and \emph{pHash-Mirror}. 

Most algorithms coped well with image scaling, likely because one of the main steps in generating perceptual hashes is to downscale the image to a manageable size/complexity.  Most algorithms have a very high exact match percentage for scaling\footnote{Curiously, the more nuanced downsampling approach taken by PDQ seems to hinder it in this case.}, peaking at 97.9\% with Wavehash, meaning that only 2.1\% of upscaled images would not have an exact hash match. Interestingly, while thumbnail performance is often almost as good, it does not produce anywhere near the same percentage of exact hash matches. All algorithms also coped very well with poor quality JPEGs.

NeuralHash produces consistently good results, though it has a very wide distribution (almost normal around 0.3) for mirroring attacks, when ideally the distribution should be very long tailed with most distances around zero.
On paper, Wavehash also falls into this category, but the false positive rate is likely too high for most use cases due to the inter-score distribution. Blockhash, pHash and PDQ all perform well on average, however, each of them have cases where the distribution is very wide due to a high standard deviation (border, crop and mirror for all three, and watermark for PDQ and pHash). Generally, mirroring an image seems to cause the most disturbance, with many algorithms essentially distributing scores as if it was an unrelated image. Borders appear to be second most impactful due to the introduction of common content.


\section{Conclusion and Future Work}
\label{subsec:conclusion}

In this work, we have explored several popular and widely used perceptual hashing algorithms in order to understand the distributions of their Hamming distances when discriminating between unrelated images, and versions of the same image. The raw data for this paper are available online with DOI: \url{10.5281/zenodo.7426035}. 
Of the tested algorithms, all but ColourHash and Wavehash were shown to have reasonable inter-score distributions, resulting in few false positives by complete chance. PDQ is particularly strong in this regard. Surprisingly, adding a border or watermark is enough to substantially throw off the distances between unrelated images for most algorithms, simply due to a small portion of shared content being introduced, biasing them towards similarity.

When comparing the images to various content-preserving modifications, mirroring the image on the x-axis was found to be particularly destructive across the board, with the addition of a bordering being a distant second. While distances should be clustered around 0 in these intra-score cases, distributions for unfavourable modifications often resulted in relatively wide spreads, sometimes centred around 0.4 and 0.5. Additionally, most techniques had at least one poorly handled image for each modification. Overall, NeuralHash appears to be the least affected by the modifications, in addition to having a desirable inter-score distribution, suggesting that deep, learned, approaches may be strong against naive user-level attack.

The understanding generated above allows for further insights to be mined in terms of setting appropriate thresholds (given some level of false positive/false negative trade-off)~\cite{mckeown_fast_2019}, but also to allow us to consider black-box mitigations against common problems in matching. Pre-processing approaches for mirrored and bordered images would help reduce negative impacts on scoring. This could be handled in a similar manner to how ForBild mitigates its algorithmic weakness to rotated images, by rotating the image such that the darkest corner is always in the same position~\cite{steinebach_robust_2011}.

The analysis above is also largely content-independent. While prior work has acknowledged that there are types of image which cause issues with certain algorithms~\cite{mckeown_fast_2019,struppek2021learning}, these particular weaknesses could be better documented at the archetype/algorithm level. This is particularly important for the impact it may have on varying corpora, such as when comparing animated images, or exported video frames using still image hashing formats.


\bibliographystyle{IEEEtran}
\bibliography{cas-refs}

\end{document}